\documentclass[]{spie}  %>>> use for US letter paper
\usepackage{amsmath,amsfonts,amssymb}
\usepackage{graphicx}
\usepackage[colorlinks=true, allcolors=blue]{hyperref}
\usepackage{float}
\usepackage{multirow}
\title{Physics-based generation of multilayer corneal OCT data via Gaussian modeling and MCML for AI-driven diagnostic and surgical guidance applications}

\author[a]{Jinglun Yu}
\author[a]{Yaning Wang}
\author[a]{Rosalinda Xiong}
\author[a]{Ziyi Huang}
\author[b,c]{Kristina Irsch}
\author[a]{Jin U.~Kang}
\affil[a]{Department~of~Electrical~and~Computer~Engineering, Johns~Hopkins~University,~Baltimore,~MD~21218,~USA}
\affil[b]
{Wilmer~Eye~Institute,
Johns~Hopkins~University,~Baltimore,~MD~21287,~USA}
\affil[c]
{Vision~Institute,~Sorbonne~University,~INSERM,~CNRS,~Paris,~France}
\authorinfo{Further author information: (Send correspondence to Jinglun Yu)\\Jinglun Yu: E-mail: jyu146@jhu.edu\\}

\begin{document} 
\maketitle
\begin{abstract}
Training deep learning models for corneal optical coherence tomography (OCT) imaging is limited by the availability of large, well-annotated datasets. We present a configurable Monte Carlo simulation framework that generates synthetic corneal B-scan optical OCT images with pixel-level five-layer segmentation labels derived directly from the simulation geometry. A five-layer corneal model with Gaussian surfaces captures curvature and thickness variability in healthy and keratoconic eyes. Each layer is assigned optical properties from the literature and light transport is simulated using Monte Carlo modeling of light transport in multi-layered tissues (MCML), while incorporating system features such as the confocal PSF and sensitivity roll-off. This approach produces over 10,000 high-resolution (1024×1024) image-label pairs and supports customization of geometry, photon count, noise, and system parameters. The resulting dataset enables systematic training, validation, and benchmarking of AI models under controlled, ground-truth conditions, providing a reproducible and scalable resource to support the development of diagnostic and surgical guidance applications in image-guided ophthalmology.
\end{abstract}

\keywords{Synthetic~OCT~data,~Monte~Carlo~modeling~of light,~OCT~corneal~imaging,~Gaussian~modeling,~AI~benchmark generation}

\section{INTRODUCTION}
\label{sec:intro} 
Corneal OCT is a cornerstone modality for diagnosing anterior-segment disease and enabling image-guided ophthalmic procedures\cite{xu2023neural,Wang2024SubretinalOCT}, but data availability remains a major bottleneck for AI. Large-scale corneal B-scan collections are hard to curate because clinical data are privacy-sensitive, acquisition protocols vary across devices, and consistent multilayer annotations are costly to produce. This limitation is especially acute in surgery-adjacent settings—such as penetrating keratoplasty (PK) and deep anterior lamellar keratoplasty (DALK)\cite{wang2024reimagining,yu2025topology,wang2023common,gensheimer2024comparison}—where layer-aware interpretation can support intraoperative decision-making and reduce uncertainty in depth-critical maneuvers.\cite{titiyal2017intraoperative}Pixel-level multilayer annotation is labor-intensive, and small datasets rarely capture the full variability of healthy and keratoconic corneas, as well as speckle and system-dependent artifacts. Data scarcity also limits the feasibility of data-hungry generative approaches (e.g., diffusion models) for augmentation or representation learning. Meanwhile, early-stage development for corneal enhancement tasks—such as super-resolution\cite{wang2025super,guo2025psi3d}, denoising, and domain adaptation—often relies on \textit{ex vivo} animal-eye data. While valuable, these data can differ from human corneal geometry and optical properties\cite{win2025corneal}, limiting their ability to support systematic robustness testing. A controllable, human-cornea–like synthetic resource could therefore provide scalable training data and enable stress-testing under known, parameterized perturbations.

Given the limited availability of large, finely annotated OCT datasets, recent work uses synthetic data in two main ways. Learning-based generators (GANs/diffusion) can expand training sets or balance classes \cite{tripathi2023generating, zheng2020assessment}, but image realism alone does not guarantee anatomically consistent layers or pixel-aligned masks, often requiring additional heuristics to derive labels \cite{win2025corneal}. Physics-based simulators provide controllable, self-consistent supervision, yet face fidelity–efficiency trade-offs: wave-equation models are costly at clinical scale, while simplified models can miss scattering and system effects \cite{brenner2019two}. Many Monte Carlo OCT simulators further prioritize A-scan settings or constrained geometries, limiting scalable high-resolution B-scan synthesis with flexible anatomy and “label-perfect” multilayer masks \cite{guo2022convolutional}.

Here we present a configurable, parallel physics-based framework that generates high-resolution synthetic corneal OCT B-scans with exact multilayer ground truth. We construct a five-layer cornea using Gaussian surface profiles to model anatomically plausible curvature and thickness variability spanning healthy and keratoconic phenotypes, and assign literature-based optical properties per layer. OCT signal formation is simulated with Monte Carlo modeling of light transport in multi-layered tissues (MCML)\cite{yao1999monte}, augmented with confocal PSF gating and sensitivity roll-off to capture key system effects under controlled perturbations\cite{wang1995mcml}. Because geometry and parameters are explicitly defined, pixel-aligned five-layer masks are exported automatically with perfect correspondence. The pipeline produces \textgreater 10,000 paired 1024×1024 image–label samples and supports systematic sweeps over curvature, thickness, photon count, noise, and PSF variants for reproducible benchmarking and robustness testing.

   \begin{figure}[H]
   \begin{center}
   \begin{tabular}{c} %% tabular useful for creating an array of images 
   \hspace{-0.2cm}
   \includegraphics[height=5.5 cm]{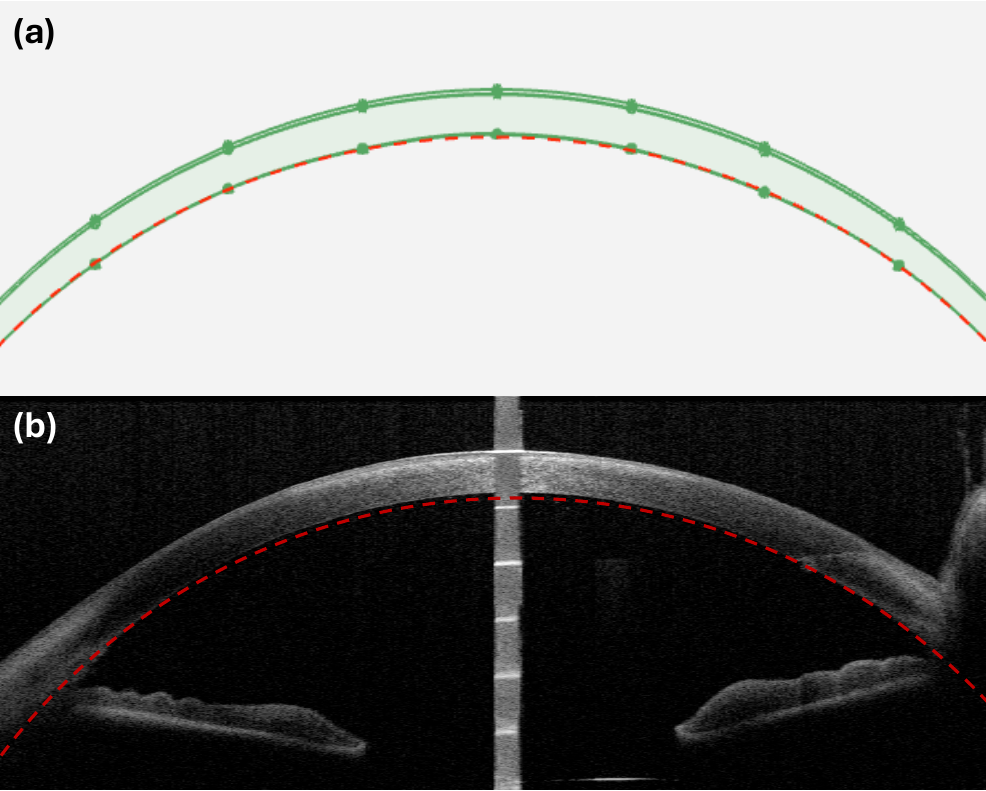}
   \end{tabular}
   \end{center}
   \caption[fig:1] 
    { \label{fig:baseline_semicircle}
    \textbf{(a)}~Semi-circular baseline curvature used as the first-order corneal support profile (red dashed curve).
    \textbf{(b)}~Clinical corneal OCT B-scan with the fitted semi-circular baseline overlaid (red dashed curve) to illustrate dome-shape consistency. Panel (b) is from the public dataset \texttt{https://doi.org/10.6084/m9.figshare.25952845}\cite{sun2024oct} and was cropped and annotated for visualization.}
   \vspace{0.1cm}
   \end{figure}
   
\section{METHODS}
\subsection{Gaussian-Based Multilayer Corneal Geometry Modeling}
\subsubsection{Boundary Parameterization and Random Sampling}
\begin{figure}[H]
   \begin{center}
   \begin{tabular}{c} %% tabular useful for creating an array of images 
   \hspace{-0.2cm}
   \includegraphics[height=5 cm]{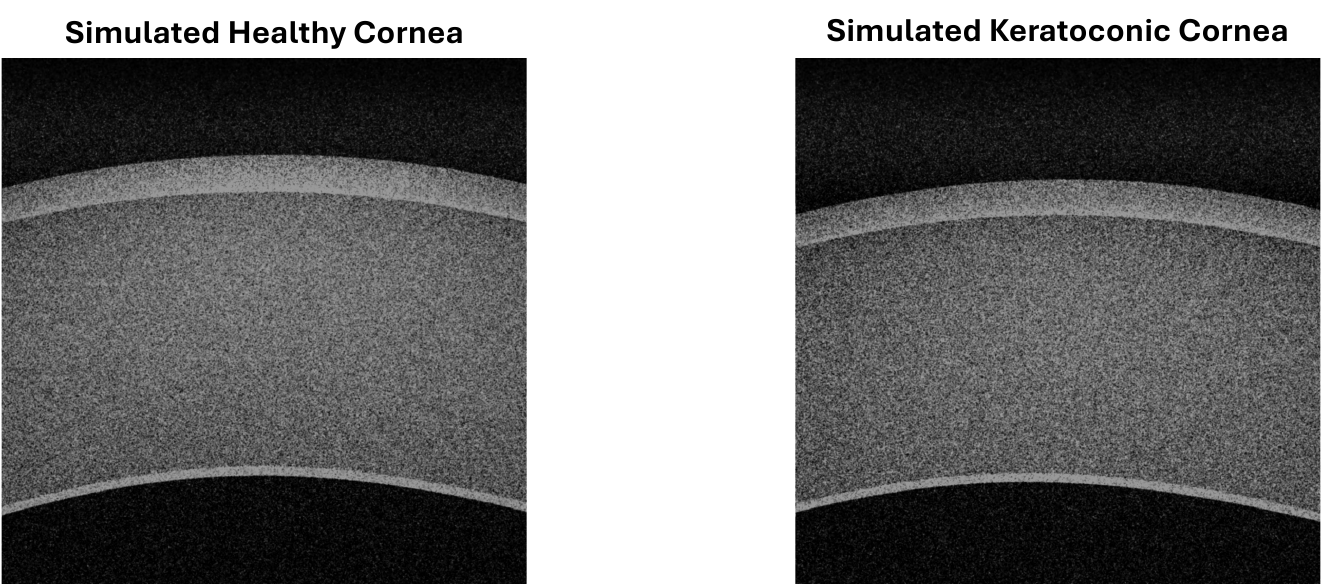}
   \end{tabular}
   \end{center}
   \caption[fig:3] 
    { \label{fig:CompareSimulation}
    Representative simulated corneal OCT B-scans generated by our model: a healthy cornea (left) and a keratoconic cornea (right), where a localized anterior steepening is introduced to mimic keratoconus-like morphology.}
   \vspace{0.1cm}
   \end{figure}

An anatomically plausible corneal B-scan geometry with controllable variability is generated by modeling each corneal layer boundary as a smooth 2D curve along the lateral axis $x$. A global baseline curvature is first defined using a semi-circular support profile,
\begin{equation}
y_{\text{base}}(x)=R-\sqrt{R^2-x^2},
\end{equation}
which captures the dome-shaped corneal curvature and provides a physically interpretable first-order fit to the dome-shaped corneal boundary and is visually validated against a representative clinical B-scan in Fig.~\ref{fig:baseline_semicircle}.
A parameterized sinusoidal--Gaussian deformation is applied to the baseline surface, enabling keratoconus-like cone formation via localized steepening (Figure~\ref{fig:CompareSimulation} shows an example of simulated healthy and keratoconus-like corneas). The deformation is defined as
\begin{equation}
\Delta y(x)=A_1\cos\!\Big(\frac{f_1\pi(x+D)}{R}\Big)
          +A_2\sin\!\Big(\frac{f_2\pi(x+D)}{R}\Big)
          +H\exp\!\Big(-\frac{(x-x_0)^2}{2\sigma^2}\Big),
\end{equation}
and the resulting boundary is given by
\begin{equation}
y(x)=y_{\text{base}}(x)+\Delta y(x).
\end{equation}
Here, $(A_1,f_1)$ and $(A_2,f_2)$ control low- and high-frequency components, $D$ introduces lateral offset (decentering), and the Gaussian term $(H,x_0,\sigma)$ controls the height, location, and width of a localized bulge. Randomness is introduced by sampling $R$, $D$, $x_0$, $(A_1,A_2,H)$, $(f_1,f_2)$, and $\sigma$ on a per-sample basis within bounded, physiologically feasible ranges (implemented as small perturbations around nominal values), enabling controlled sweeps over anatomical variability while maintaining smooth and realistic curvature. A fixed random seed can optionally be used to reproduce identical geometries across runs.

\subsubsection{Multilayer Model Construction and Calibration}
\begin{figure}[H]
   \begin{center}
   \begin{tabular}{c} %% tabular useful for creating an array of images 
   \hspace{-0.2cm}
   \includegraphics[height=5 cm]{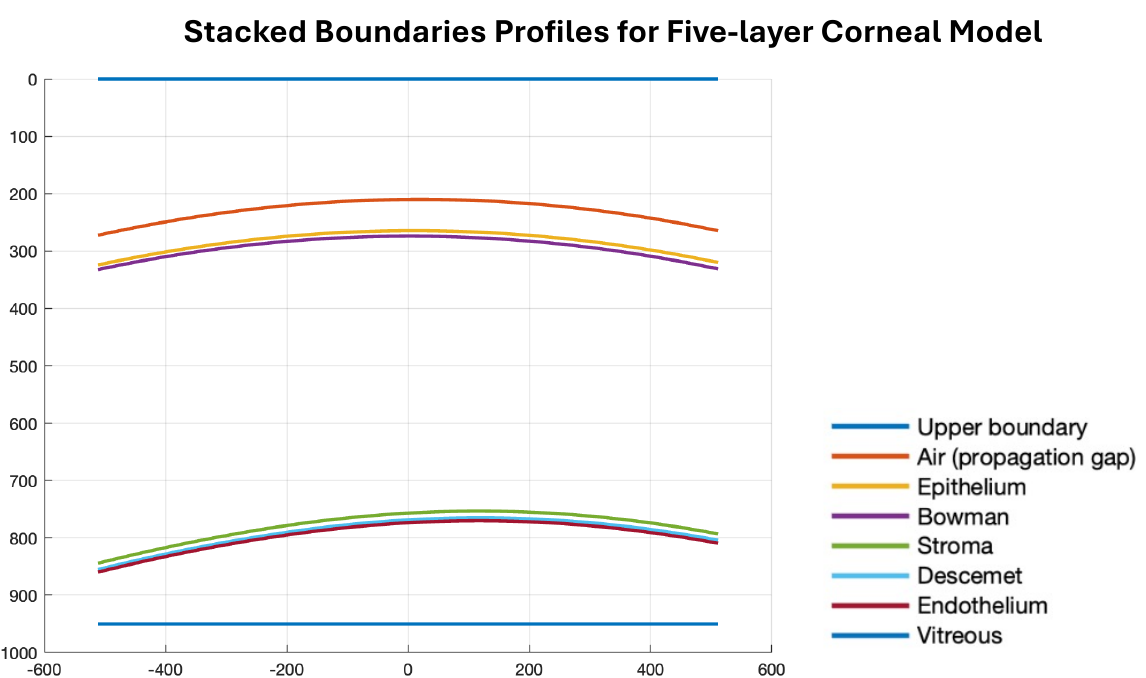}
   \end{tabular}
   \end{center}
   \caption[fig:3] 
    { \label{fig:depthinformation}Stacked boundaries profiles used to construct a five-layer corneal model. Each interface curve defines one layer boundary, enabling independent control of the thickness profiles for the epithelium, Bowman’s layer, stroma, Descemet’s membrane, and endothelium; the upper boundary, air (propagation gap), and vitreous boundary are included for completeness.}
   \vspace{0.1cm}
   \end{figure}

A five-layer cornea (epithelium, Bowman’s layer, stroma, Descemet’s membrane, and endothelium) is constructed by stacking boundaries with independently controlled thickness profiles, as illustrated in Fig.~\ref{fig:depthinformation}. Given an initial boundary $y_0(x)$, subsequent interfaces are generated as
\begin{equation}
y_{k+1}(x)=y_k(x)+t_k(x),
\end{equation}
where $t_k(x)$ denotes the thickness of layer $k$. Inter-sample variability is introduced by applying a unified multiplicative perturbation to nominal profiles,
\begin{equation}
t_k(x)=s\,t_k^{(0)}(x), \qquad s\sim \mathcal{U}(1-\delta,\,1+\delta), \;\; \delta=0.05\text{--}0.10,
\end{equation}
followed by adaptive constraint enforcement: if any boundary overlap or ordering violation occurs, the perturbation is damped (i.e., $s$ is reduced) until all interfaces remain non-intersecting and layer-specific lower bounds are satisfied. 

\begin{figure}[H]
   \begin{center}
   \begin{tabular}{c} %% tabular useful for creating an array of images 
   \hspace{-0.2cm}
   \includegraphics[height=5 cm]{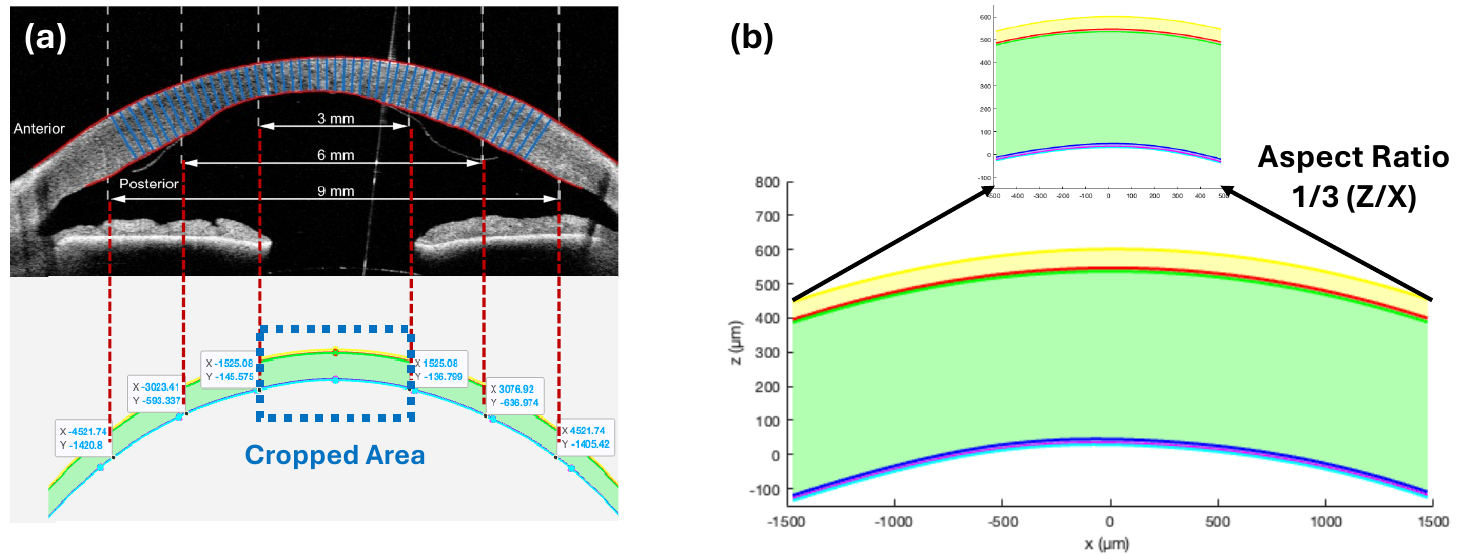}
   \end{tabular}
   \end{center}
   \caption[fig:3] 
    { \label{fig:aspect_ratio_calibration}
    \textbf{(a)} Simulated cornea section anchored to clinically relevant scan widths (3--9~mm). The upper portion of panel (a) includes adapted material from Scientific Reports (DOI: \texttt{10.1038/s41598-021-93186-9}) under CC BY 4.0\cite{heslinga2021corneal} and was re-rendered and annotated.
    \textbf{(b)}~Cropped multilayer corneal area from panel (a) and the final calibrated visualization, obtained by applying the aspect-ratio.}
   \vspace{0.1cm}
   \end{figure}

To match clinical B-scan metrics and standard OCT visualization, the geometry is rendered with a calibrated lateral--axial aspect ratio. As shown in Fig.~\ref{fig:aspect_ratio_calibration}, the multilayer region is then cropped to a central, clinically relevant window before forward simulation and label rasterization and an axial scaling relative to the lateral axis ($Z/X \approx 1/3$) is applied to preserve clinically meaningful curvature and layer spacing, anchored to representative scan widths. 

\subsection{Physics-Based OCT Forward Simulation via MCML}

\subsubsection{Optical Coefficient Specification and Map Generation}
Layer-wise optical properties are specified under a piecewise-homogeneous assumption, where each anatomical layer is assigned constant refractive index $n$, absorption coefficient $\mu_a$, scattering coefficient $\mu_s$, and anisotropy factor $g$ based on literature-reported values~\cite{wang2004objective, boettner1962transmission, meek2015corneal, tuchin2015tissue, bashkatov2018measurement}. The coefficients used in this work are summarized in Table~\ref{tab:optical_coeffs}. The column $d$ reports the nominal (average) layer thickness in centimeters for reference.

\begin{table}[t]
\centering
\caption{Layer-wise optical properties for MCML forward simulation (literature-based constants). 
The tear film (thickness $\approx 3\,\mu$m) is omitted and modeled implicitly via the air--epithelium interface. 
$d$ is the nominal average layer thickness (cm) used in simulation.}
\label{tab:optical_coeffs}
\begin{tabular}{lccccc}
\hline
Layer & $n$ & $\mu_a$ (cm$^{-1}$) & $\mu_s$ (cm$^{-1}$) & $g$ & $d$ (cm) \\
\hline
Air (propagation gap) & 1.000 & 0.00 & 0.00  & 0.00 & 0.0262 \\
Epithelium     & 1.400 & 0.20 & 10.00 & 0.92 & 0.0052 \\
Bowman         & 1.400 & 0.20 & 8.00  & 0.92 & 8.3612e-04 \\
Stroma         & 1.376 & 0.20 & 4.50  & 0.94 & 0.0489 \\
Descemet       & 1.375 & 0.30 & 8.00  & 0.93 & 0.0011 \\
Endothelium    & 1.375 & 0.30 & 10.00 & 0.93 & 4.4537e-04 \\
Vitreous Body  & 1.336 & 0.06 & 0.02  & 0.93 & 0.0124 \\
\hline
\end{tabular}
\end{table}

For each simulated B-scan, the multilayer geometry is represented by ordered interfaces $\{y_k(x)\}_{k=1}^{K}$ (in $\mu$m) along the lateral coordinate $x$, where $K=6$ corresponds to the six corneal boundaries and an additional bottom limit closes the simulation window. From these interfaces, a per-A-line layer-thickness map is computed via adjacent boundary differences,
\begin{equation}
\mathbf{t}(x)=\big[t_{\mathrm{air}}(x),\,t_{\mathrm{epi}}(x),\,t_{\mathrm{bow}}(x),\,t_{\mathrm{str}}(x),\,t_{\mathrm{des}}(x),\,t_{\mathrm{end}}(x),\,t_{\mathrm{vit}}(x)\big]^{\top},
\end{equation}
with, for example, $t_{\mathrm{epi}}(x)=y_{\mathrm{air}}(x)-y_{\mathrm{epi}}(x)$ and $t_{\mathrm{bow}}(x)=y_{\mathrm{epi}}(x)-y_{\mathrm{bow}}(x)$. Thickness profiles are converted to centimeters for MCML input\cite{guo2022convolutional}.

A pixel-wise layer index map $L(x,z)$ is then derived by assigning each pixel at axial coordinate $z$ (increasing downward in the rendered B-scan) to the unique layer interval bounded by adjacent interfaces:
\begin{equation}
L(x,z)=\ell \quad \text{if} \quad y_{\ell-1}(x) \ge z > y_{\ell}(x),
\end{equation}
where $\ell\in\{1,\dots,7\}$ denotes the layer (Air, Epithelium, \dots, Vitreous), and $y_0(x)$ and $y_7(x)$ denote the top and bottom limits of the cropped window, respectively. Given $L(x,z)$, coefficient maps are generated by piecewise-constant projection:
\begin{equation}
n(x,z)=n_{L(x,z)},\quad
\mu_a(x,z)=\mu_{a,L(x,z)},\quad
\mu_s(x,z)=\mu_{s,L(x,z)},\quad
g(x,z)=g_{L(x,z)},
\end{equation}
yielding pixel-aligned optical property maps exactly consistent with the sampled multilayer geometry.

\subsubsection{MCML with Sensitivity Roll-off and Confocal Gating}
The OCT forward model is implemented using the Monte Carlo simulation of photon transport in a multilayer scattering medium parameterized by $(\mu_a,\mu_s,g,n)$ per layer\cite{wang1995mcml}. Photon step lengths are sampled from an exponential distribution,
\begin{equation}
s=-\frac{\ln(\xi)}{\mu_t}, \qquad \mu_t=\mu_a+\mu_s,\;\; \xi\sim\mathcal{U}(0,1),
\end{equation}
with photon weight reduced according to the absorption fraction $\mu_a/\mu_t$ at each interaction. Scattering directions are sampled from the Henyey--Greenstein phase function
\begin{equation}
f_{\mathrm{HG}}(\cos\theta)=\frac{1-g^2}{2\left(1+g^2-2g\cos\theta\right)^{3/2}},
\end{equation}
using inverse-CDF sampling for $\theta$ and uniform sampling for the azimuthal angle. Boundary crossings are handled using Fresnel reflection/refraction based on refractive indices, and low-weight photons are terminated via Russian roulette. Detected photons are accumulated into a depth-resolved A-line $R(z,x)$\cite{wang1995mcml}.

To approximate system-dependent OCT sensitivity, a parametric, implementation-level model is applied via depth-dependent confocal gating and sensitivity roll-off. Confocal weighting is modeled as a Gaussian envelope centered at the focal depth $z_0$,
\begin{equation}
W_{\mathrm{conf}}(z)=\exp\!\left(-\frac{(z-z_0)^2}{2\sigma_c^2}\right),
\end{equation}
and sensitivity roll-off is modeled as
\begin{equation}
W_{\mathrm{roll}}(z)=\cos^4\!\left(\frac{\pi z}{z_{\max}}\right),
\end{equation}
chosen to emulate typical depth-dependent sensitivity decay while providing a tunable attenuation profile. As implemented here, the roll-off can be further down-weighted beyond mid-range,
\begin{equation}
W_{\mathrm{roll}}(z)\leftarrow \eta\,W_{\mathrm{roll}}(z)\quad \text{for}\quad z>\frac{z_{\max}}{2}, \qquad \eta\in(0,1),
\end{equation}
and the final system-weighted OCT signal is computed by
\begin{equation}
R_{\mathrm{sys}}(z,x)=R(z,x)\,W_{\mathrm{conf}}(z)\,W_{\mathrm{roll}}(z),
\end{equation}
with a small floor $R_{\mathrm{sys}}(z,x)\leftarrow \max(R_{\mathrm{sys}}(z,x),w_{\min})$ to avoid complete signal collapse at large depths. The displayed B-scan is obtained by log compression and contrast normalization for visualization.

\section{Experiments and Results}

\subsection{Dataset Composition and Experimental Protocol}
\begin{figure}[H]
  \begin{center}
  \begin{tabular}{c}
  \hspace{-0.2cm}
  \includegraphics[height=7.5cm]{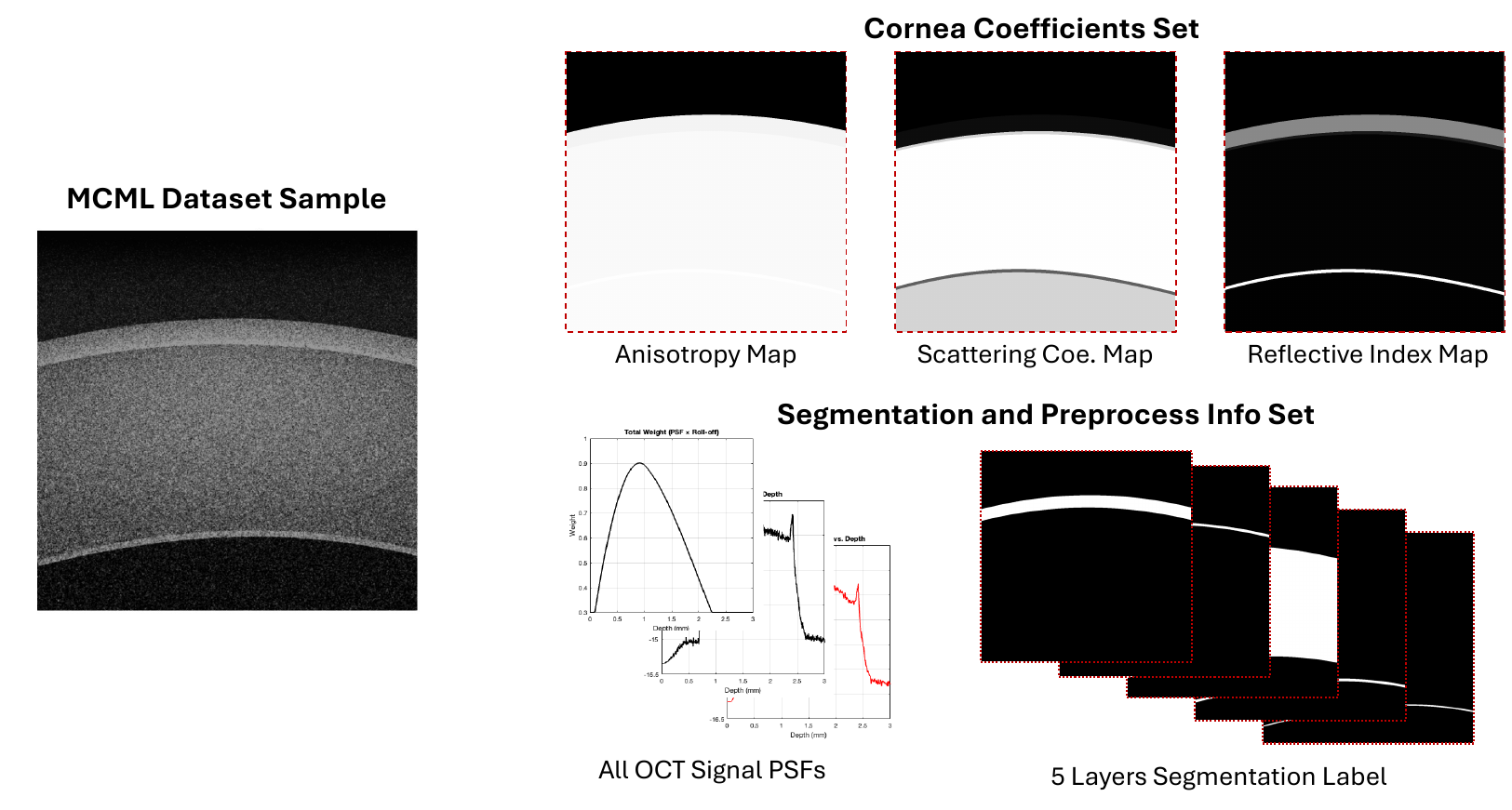}
  \end{tabular}
  \end{center}
  \caption[Per-sample outputs of the synthetic dataset]{
  \label{fig:dataset}A representative sample set produced by the proposed framework.}
  \vspace{0.1cm}
\end{figure}

The proposed framework generates a synthetic corneal OCT dataset of 10,000 B-scans, comprising 8,000 healthy samples and 2,000 keratoconus-like samples produced via controlled geometry and optical-parameter perturbations. As illustrated in Fig.~\ref{fig:dataset}, each sample is packaged as a paired set including a simulated OCT B-scan, pixel-accurate five-layer segmentation masks, three optical-parameter maps (refractive index $n$, scattering coefficient $\mu_s$, and anisotropy factor $g$), and the corresponding system-effect--modulated OCT signals (e.g., confocal/PSF gating and sensitivity roll-off). The experiments in this work are conducted on the healthy subset to establish a clean benchmarking protocol. For computational efficiency, both tasks use randomly sampled B-scans from the full dataset (healthy + keratoconus): reconstruction is trained/tested on 1{,}000/500 samples, and segmentation on 1{,}000/100 samples.

\subsection{Task I: Diffusion Benchmarking for Joint Optical-Map and Structural OCT Reconstruction}
\begin{figure}[t]
  \centering
  \includegraphics[height=6cm]{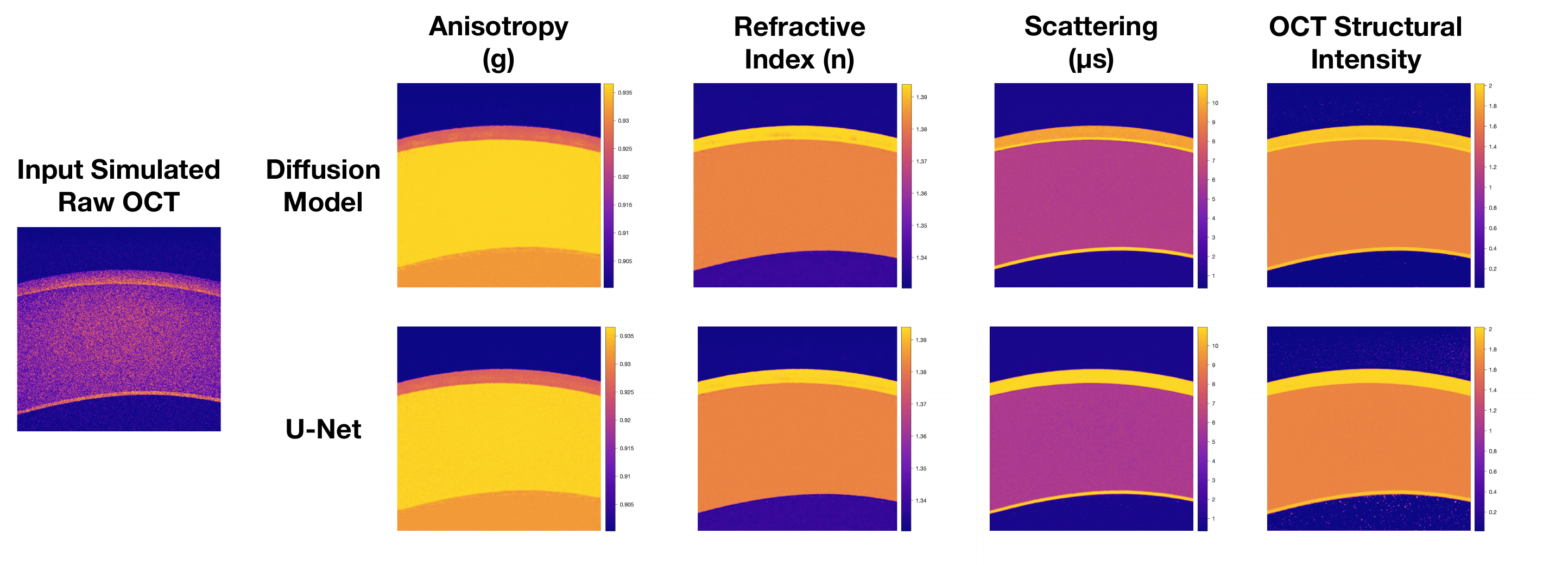}
  \caption[Qualitative comparison for Task I]{
  \label{fig:diffusionresult}Qualitative comparison on a representative healthy sample for Task I.}
  \vspace{0.1cm}
\end{figure}

This experiment evaluates whether the proposed synthetic dataset can serve as a controlled benchmark for diffusion-based inverse OCT reconstruction under paired ground truth. Using healthy samples, the model takes a simulated raw OCT B-scan as input and jointly predicts three optical-parameter maps $(n,\mu_s,g)$ together with a structural OCT intensity output. A physics-based forward model is used during training to encourage signal-consistency between the predicted optical maps and the reconstructed OCT intensity. Performance is quantified using PSNR/SSIM/MSE for both the structural OCT output and the recovered parameter maps. Table~\ref{tab:task1_full_vs_unet} reports the quantitative comparison between the diffusion-based full model and a baseline U-Net. The diffusion model achieves markedly higher fidelity for structural OCT reconstruction and improved recovery of scattering-related parameters ($\mu_s$ and $g$), while the baseline remains competitive for refractive-index reconstruction. Figure~\ref{fig:diffusionresult} provides a representative qualitative comparison, where the diffusion model yields cleaner layer separation and more consistent parameter contrast across the corneal interfaces.

\begin{table}[h]
\centering
\caption{Task I results on healthy data for joint reconstruction. The Diffusion Model denotes the diffusion-based reconstruction approach, illustrating the dataset utility for diffusion benchmarking under paired ground truth.}
\label{tab:task1_full_vs_unet}
\setlength{\tabcolsep}{5pt}
\renewcommand{\arraystretch}{1.05}
\begin{tabular}{l|ccc|ccc}
\hline
\multirow{2}{*}{Method} 
& \multicolumn{3}{c|}{OCT} 
& \multicolumn{3}{c}{$n$} \\
& PSNR$\uparrow$ & SSIM$\uparrow$ & MSE$\downarrow$
& PSNR$\uparrow$ & SSIM$\uparrow$ & MSE$\downarrow$ \\
\hline
Diffusion Model 
& 31.65 & 0.94 & 2.75e-3
& 28.76 & 0.66 & 6.54e-6 \\
Baseline U-Net 
& 25.04 & 0.85 & 1.2e-2
& 29.29 & 0.67 & 5.78e-6 \\
\hline
\multicolumn{7}{c}{}\\[-1.0ex]
\hline
\multirow{2}{*}{Method} 
& \multicolumn{3}{c|}{$\mu_s$} 
& \multicolumn{3}{c}{$g$} \\
& PSNR$\uparrow$ & SSIM$\uparrow$ & MSE$\downarrow$
& PSNR$\uparrow$ & SSIM$\uparrow$ & MSE$\downarrow$ \\
\hline
Diffusion Model
& 28.03 & 0.70 & 0.23
& 29.25 & 0.84 & 1.90e-6 \\
Baseline U-Net 
& 24.09 & 0.63 & 0.56
& 27.89 & 0.78 & 2.60e-6 \\
\hline
\end{tabular}
\end{table}

\subsection{Task II: Baseline Evaluation for Three-Class Corneal Layer Segmentation}
This experiment evaluates a simple segmentation baseline on the proposed synthetic corneal OCT dataset with pixel-accurate labels. To reduce task complexity and emphasize benchmarking utility, the original five corneal layers are merged into three classes: Epithelium (EPI), Bowman+Stroma (BS), and Descemet+Endothelium (DE). A vanilla U-Net is trained on healthy samples with a binary cross-entropy loss with logits and Adam optimization (learning rate $10^{-4}$) for 200 epochs, using a random 9:1 train/test split.

\begin{figure}[h]
  \centering
  \includegraphics[width= 14 cm]{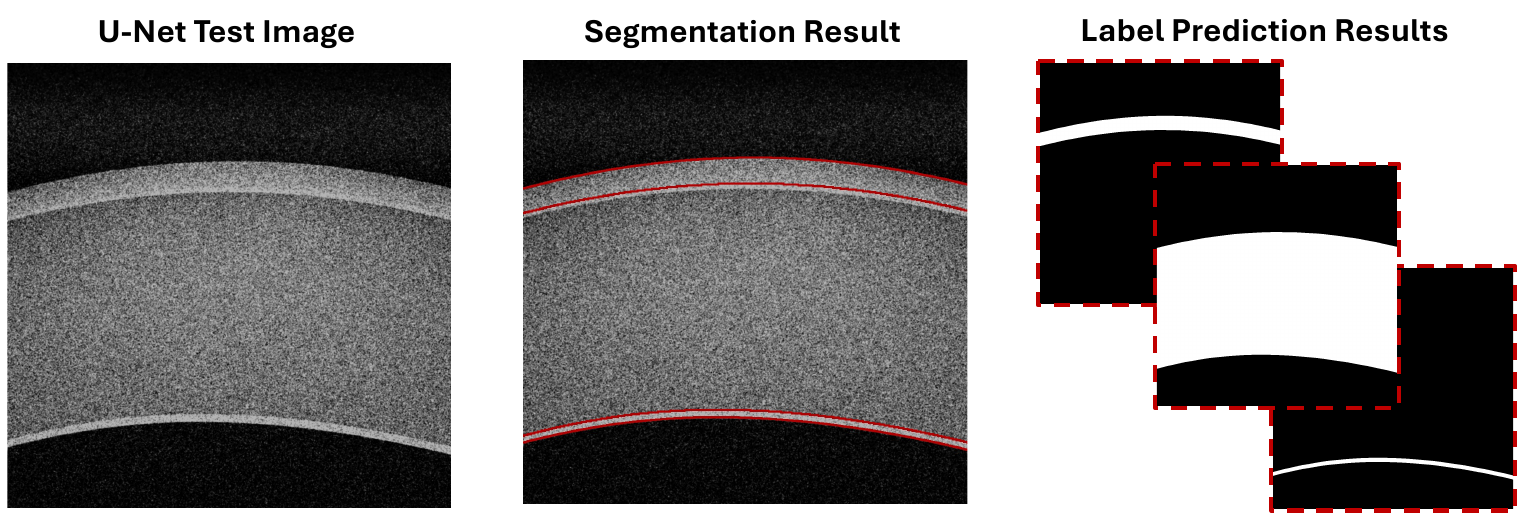}
  \caption[Qualitative segmentation assessment for Task II]{
  \label{fig:Unetresult}
  Qualitative assessment of three-class layer segmentation.}
  \vspace{0.1cm}
\end{figure}

Segmentation performance is quantified by per-class Intersection-over-Union (IoU) and mean squared error (MSE) on the held-out test set. As summarized in Table~\ref{tab:task2_seg_mean}, the baseline achieves near-ceiling accuracy for BS and EPI, while DE remains comparatively more challenging. For qualitative assessment, as Fig.~\ref{fig:Unetresult} shows, boundary curves are extracted from predicted masks by identifying per-column upper/lower interfaces and rendered as overlaid contours on the input OCT B-scan, enabling direct inspection of interface alignment and error patterns.

\begin{table}[t]
\centering
\caption{Task II mean segmentation performance on the held-out healthy test set (three merged classes).}
\label{tab:task2_seg_mean}
\setlength{\tabcolsep}{8pt}
\renewcommand{\arraystretch}{1.1}
\begin{tabular}{lcc}
\hline
Class & IoU$\uparrow$ & MSE$\downarrow$ \\
\hline
Bowman+Stroma (BS)          & 0.998865 & 5.99$\times 10^{-4}$ \\
Descemet+Endothelium (DE)  & 0.980385 & 3.35$\times 10^{-4}$ \\
Epithelium (EPI)           & 0.992192 & 4.50$\times 10^{-4}$ \\
\hline
\end{tabular}
\end{table}

\section{DISCUSSION AND CONCLUSION}
The proposed method introduces a configurable physics-based pipeline for generating high-resolution synthetic corneal OCT B-scans with label-perfect supervision, directly addressing the data scarcity that limits training and stress-testing modern AI models. By combining Gaussian-parameterized multilayer geometry modeling with MCML and key system effects (confocal gating and sensitivity roll-off), the framework outputs paired OCT images, pixel-accurate five-layer masks, and aligned optical-parameter maps. The resulting dataset comprises 10,000 samples spanning healthy and keratoconus-like phenotypes, providing scalable data enlargement and controlled variation for reproducible training, ablation, and robustness evaluation.

Benchmarking on the healthy subset demonstrates two utilities for AI development: (i) in Task~I, the diffusion-based reconstruction model substantially improves structural OCT fidelity and scattering-related parameter recovery over a baseline U-Net, showing that the dataset supports data-hungry generative inverse models with paired ground truth; and (ii) in Task~II, a vanilla U-Net achieves near-ceiling IoU on merged three-class segmentation, validating geometric/label consistency and enabling fast, standardized baseline comparisons. Remaining gaps include simplified layer-wise optics and parametric system weighting; future work will incorporate richer scanner/noise effects and evaluate domain shift using the keratoconus subset to improve transfer to clinical OCT.

\acknowledgments % equivalent to \section*{ACKNOWLEDGMENTS}       
 This work was supported by National Institute of Biomedical
Imaging and Bioengineering of the National Institutes of Health Grant under award number 1R01EY032127 (PI: Jin U.~Kang). The study was conducted at Johns~Hopkins~University.

% References
\bibliography{report} % bibliography data in report.bib
\bibliographystyle{spiebib} % makes bibtex use spiebib.bst
   
\end{document}